\documentstyle[aps,prl,twocolumn,psfig]{revtex}

\def\Bm{{B}}

\def\Em{\E}
\def\E{{\cal E}}
\def\Hm{H}
\def\H{{\cal H}}
\def\MC{Monte Carlo}
\def\Nc{N_{\rm c}}
\def\Nm{N}

\def\beq{\begin{equation}}
\def\betah{\hat\beta}

\def\cfMC{correlation function \MC}
\def\d #1 #2{d_{#1}^{(#2)}}
\def\dMC{diffusion \MC}
\def\eeq{\end{equation}}
\def\etal{{et al.}}

\def\hr{\hat r}

\def\psig{\psi_{\rm g}}

\def\tE{\tilde E}
\def\tpsi #1{\tilde{\psi}^{(#1)}}
\def\tpsih #1{\hat{\psi}^{(#1)}}

\def\vv{{v.v.}}

\begin{document}
\setlength{\topmargin}{0.3cm}
\title{Optimization of ground and excited state wavefunctions and van
  der Waals clusters}
\author{M. P.  Nightingale and Vilen Melik-Alaverdian}
\address{ Department of Physics, University of Rhode Island, Kingston
  RI 02881, USA} 
\date{\today} 
\maketitle

\begin{abstract}
  A quantum \MC\ method is introduced to optimize excited state trial
  wavefunctions.  The method is applied in a correlation function
  \MC\ calculation to compute ground and excited state energies of
  bosonic van der Waals clusters of upto seven particles.  The
  calculations are performed using trial wavefunctions with general
  three-body correlations.

\end{abstract}

Weakly-bound clusters display strong anharmonicity, and this makes
solving the Schr\"odinger equation for such systems a formidable
computational challenge.  The discrete variable representation method
(DVR)\cite{ZWZvDB95} has been applied with success to compute the
energies of vibrational states for systems with upto six degrees of
freedom.  Its computational complexity scales exponentially with
dimension, a problem that \MC\ methods can avoid.  Indeed,
\cfMC\cite{CeperleyBernu88,moreCFMC} and the projector operator
imaginary time spectral evolution (POITSE)\cite{BWjcp00} methods are
applicable to higher dimensional systems, although in practice they are
restricted to a smaller number of excited states.

The accuracy of \MC\ projection methods can be improved
dramatically by employing approximate eigenfunctions. In fact, without
good initial approximations one rarely obtains results of sufficient
accuracy.  In this Letter, we discuss a systematic and efficient
method to construct approximate eigenfunctions by optimization of
many-parameter trial functions.  We then use these functions in a
\cfMC\ calculation.  We expect that also POITSE calculations can be
improved by the same means.  A variant of the method described here
was applied previously to study critical dynamics of lattice
systems.\cite{NBdyn.prl98,NBprb00}

We compute energy levels of bosonic van der Waals clusters of atoms of
mass $\mu$, interacting via a Lennard-Jones potential with core radius
$\sigma$ and well depth $\epsilon$.  In dimensionless form, the pair
potential is $r^{-12}-2r^{-6}$ and the Hamiltonian is $\H=P^2/2m+V$;
$P^2/2m$ and $V$ are the total kinetic and potential energy.  The only
parameter is the inverse dimensionless mass $m^{-1}=\hbar^2 /\mu
\sigma^2 \epsilon$, which is proportional to the square of the de Boer
parameter.\cite{deBoer}

We use the position representation, and denote by $R$ the Cartesian
coordinates of a cluster of $\Nc$ atoms.  For the parameter
optimization we generate a sample of configurations $R_\sigma$,
$\sigma=1,\dots,\Sigma$, sampled with relative probabilities
$\psig(R_\sigma)^2$; the choice of the guiding function $\psig^2$ will
be discussed later.  The sample typically consists of several thousands
configurations, which are kept fixed during the optimization.

The trial functions are linear combinations of about a hundred {\it
  elementary basis functions,} each of which depends on non-linear
parameters. Correspondingly, we have a linear optimization nested in a
non-linear one.  The result is a set of functions serving as
{\it basis functions} in a \cfMC\ calculation.  These basis functions
are constructed one at a time, from the ground state up, as follows.

Suppose we fix at initial values the non-linear parameters of the
elementary basis functions denoted by $\beta_i$, $i=1,\dots,n$.
Ideally, these functions span an $n$-dimensional invariant subspace of
the Hamiltonian $\H$, and then there exists an $n\times n$ matrix $\E$
so that
\begin{equation}
\H \beta_i(R_\sigma)=\sum_j \beta_j(R_\sigma) \E_{ji}.
\label{eq.Hb}
\end{equation}
In that case, for $k=1,\dots,n$,
\begin{equation}
\tpsi k (R)=\sum_i \beta_i(R)\,\d i k
\label{eq.tpsik}
\end{equation}
is an eigenvector of $\H$ with an eigenvalue $\tE_k$ equal to the
exact energy $E_k$, if $d^{(k)}$ is a right eigenvector of $\E$ with
eigenvalue $\tE_k$.  We rewrite Eq.~(\ref{eq.Hb}) in matrix form
\begin{equation}
\Bm^\prime=\Bm\Em,
\label{eq.B'}
\end{equation}
where $B_{\sigma i} = \betah_i(R_\sigma)$ and $B^\prime_{\sigma
  i}=\betah^\prime_{i}(R_\sigma)$, with $\betah_i=\beta_i/\psig$ and
$\betah^\prime_i=\H\beta_i/\psig$.

In practice, the subspace spanned by the basis functions is not
invariant, so that for given matrices $\Bm$ and $\Bm^\prime$
Eq.~(\ref{eq.B'}) is an overdetermined set of equations for the
unknown matrix $\Em$.  If one multiplies through from the left by
$\Bm^T$, the transpose of $\Bm$, one obtains by inversion
\begin{equation}
\Em=(\Bm^T\Bm)^{-1}(\Bm^T\Bm^\prime)\equiv\Nm^{-1}\Hm.
\label{eq.NinvH}
\end{equation}
As readily verified, this is the least-squares solution of
Eq.~(\ref{eq.B'}); note that the rows of the matrices $\Bm$ and
$\Bm^\prime$ are weighted by the guiding function so that the elements
of the matrices $\Nm$ and $\Hm$ approach the standard quantum
mechanical overlap integrals and matrix elements in the limit of an
infinite \MC\ sample.  Eqs.~(\ref{eq.tpsik}) and (\ref{eq.NinvH}) are
usually derived from stationarity (in the linear parameters) of the
average energy.  If the latter is estimated by a finite-sample
average, requiring stationarity of this estimate yields
Eq.~(\ref{eq.NinvH}) with $H$ replaced by its {\it symmetrized}
analog.  Since the exact quantum mechanical expression is indeed
symmetrical, one might be inclined to use the symmetrized $H$.
However, only the non-symmetric expression $\Bm^T\Bm^\prime$ in
Eq.~(\ref{eq.NinvH}) satisfies the zero-variance principle of yielding
exact results independent of the configuration sample if the basis
functions span an invariant subspace of the Hamiltonian.  As in the
ideal case, Eq. (\ref{eq.tpsik}) determines the linearly optimized
trial functions, but now one has $E_k \alt \tE_k$, an
inequality\cite{MacDonald} which for a finite \MC\ sample may be
violated because of statistical noise.

The solution for $\Em$ as written in Eq.~(\ref{eq.NinvH}) is
numerically unstable since the matrix $N$ is ill conditioned because
of near-linear dependence of the $\beta_k$.  The solution to this
problem\cite{GolubVanLoan,NumericalRecipes.SVD} is to use a singular
value decomposition to obtain a numerically regularized inverse
$\Bm^{-1}$\cite{close.to.0}.  In terms of the latter, one finds from
Eq.~(\ref{eq.B'})
\begin{equation}
\Em=\Bm^{-1}\Bm^\prime.
\label{eq.Esvd}
\end{equation}

With the linear variational parameters optimized for fixed non-linear
parameters in the elementary basis functions, we optimize --following
Umrigar \etal\cite{CyrusOptimization}-- the non-linear parameters by
minimization of
\begin{equation}
\chi^2=
{
{\sum_\sigma \left({\tpsih k}{^\prime(R_\sigma)}-\tE_k \tpsih k (R_\sigma) \right)^2}
\over
{\sum_\sigma \tpsih k (R_\sigma)^2}},
\label{eq.chi2}
\end{equation}
the variance of the local energy of the wavefunction given in
Eq.~(\ref{eq.tpsik}); again, the guiding function is incorporated via
$\tpsih k=\tpsi k/\psig$ and ${\tpsih k}{^\prime}= \H\tpsi k/\psig$.

We now address the choice of the guiding function $\psig$.  To obtain
acceptable statistical errors, the sample has to have sufficient
overlap with the desired excited states.  In our case, this can be
accomplished\cite{CeperleyBernu88} with $\psig^{\;p} = \tpsi 1$ with a
power $p$ in the range $2 \alt p \alt 3$, while the ground state
wavefunction $\tpsi 1$ is obtained after a few initial iterations.

The elementary basis functions\cite{MN94,MMN96} are the final
ingredient of the computation. They are defined as functions of all
interatomic distances $r_{\sigma\tau}$ and scaled variables
$\hr_{\sigma\tau}=f(r_{\sigma\tau})$. Here, $f$ maps the interatomic
distances monotonically onto the interval $(-1,1)$ such that most of
the variation occurs where the wavefunction differs most from zero;
the explicit form of $f$ is not important for the current
discussion.

The elementary basis functions used for energy level $k$ have
non-linear variational parameters $a^{(k)}_j$, and are of the form
\begin{eqnarray}
& &\beta_i(R)=\nonumber\\
& &\ \ s_i(R)\exp\left({\sum_j a^{(k)}_j s_j(R)-
\sum_{\sigma<\tau} \left(\kappa_k r_{\sigma\tau}+
{\sqrt m\over 5 r_{\sigma\tau}^5}\right)}\right);
\label{eq.betai}
\end{eqnarray}
$\sigma,\tau,\upsilon=1,\dots,\Nc$ are atom indices.  The
polynomial $s_i$ is characterized by three non-negative integral
powers $n_{il}$:
\begin{equation}
s_i(R)=\sum_{\sigma<\tau<\upsilon} \prod_{l=1}^3 
(\hr_{\sigma\tau}^l+\hr_{\tau\upsilon}^l+
\hr_{\upsilon\sigma}^l)^{n_{il}}.
\end{equation}
The prefactor polynomial $s_i$ has bosonic symmetry, and contains
general three-body correlations, since all polynomials symmetric in
$x,y$, and $z$ can be written as polynomials in the three invariants
$I_l=x^l+y^l+z^l$, with $l=1,2,3$ and \vv\cite{MN94} The number of
elementary basis functions is limited by a bound on the total degree
$\sum_l ln_{il}$; the polynomials $s_j$ in the exponent are of the
same form as those in the prefactor, and their number in
Eq.~(\ref{eq.betai}) is limited similarly.

The constant $\kappa_k$ is determined self-consistently so that the
wavefunction has the correct exponential decay in the limit that a
single atom goes off to infinity.  Assuming --as is plausible for the
small clusters studied here-- that the energy of a cluster is roughly
proportional to the number of atom pairs\cite{Leitner91}, we find
\begin{equation}
\kappa_k={2 \over \Nc-1}\sqrt{-m\tE_k\over \Nc}.
\end{equation}
The $r_{\sigma\tau}^{-5}$ term in Eq.~(\ref{eq.betai}) ensures that
$\H\beta_i/\beta_i$ has a weaker divergence than with
$r_{\sigma\tau}^{-12}$ in the limit $r_{\sigma\tau}\to 0$.\cite{MN94}

States of higher energy are found with the same optimization scheme by
using the appropriate eigenvector $d^{(k)}$ of the matrix $\E$ in
Eq.~(\ref{eq.tpsik}).  We use the same scaling function $f$ for all
states, but different non-linear parameters $a^{(k)}_j$ and
$\kappa_k$.  This scheme works as long as the trial functions possess
the variational freedom accurately to represent the eigenstates.
Otherwise, for the \MC\ samples of the size we are using, states may be
skipped or spuriously introduced.  We found it useful to check
consistency of eigensystems obtained with this basis set and one that
includes the variational wavefunctions of previously determined,
lower-lying states.

Instead of the hybrid method discussed in this Letter, which treats
linear and non-linear variational parameters differently, we originally
attempted to use minimization of the variance of the energy for both
types of parameters simultaneously.  Although that works for
statistical mechanical applications,\cite{NBprb00} it fails here,
except for the lowest two or three energy levels.  We could only make
this work (and even then only for small de Boer parameters) for higher
levels by first constructing approximate wave functions using a
conventional approximation, and then fitting these functions by the
basis functions used in this Letter.  Finally, the parameter values
obtained in these fits served as starting values for further
optimization.\cite{MNijmp00}

We used the optimized trial wavefunctions as basis functions in a
correlation function \MC\ calculation.\cite{CeperleyBernu88,moreCFMC}.
Formally, this means that the $n$ elementary basis functions $\beta_i$
are replaced by a small number of functions $\exp(-{1 \over
  2}t\H)\tpsi k$.  For this part of the computation, the analog of
Eq.~(\ref{eq.NinvH}) is used to compute eigenvalues, rather than
Eq.~(\ref{eq.Esvd}).  The singular value decomposition (SVD) cannot be
used for the \cfMC\ because there are too many configurations
$R_\alpha$ to store the required matrices $B$ and $B^\prime$.  However,
since the optimized basis functions are few in number and roughly
orthonormal --at least for small projection times $t$-- the SVD is not
essential in this case.

Before we present estimates of the excited state energies, we discuss
the sources of error of this method.\cite{CeperleyBernu88} In addition
to the statistical errors, there are two systematic errors.  For any
finite projection time $t$ and in the limit of vanishing statistical
errors, the energies computed by this method are upper bounds to the
exact energies.\cite{MacDonald} In practice, since the statistical
errors increase with projection time, one should choose the smallest
projection time such that the projection and statistical errors are of
the same order of magnitude.  To pinpoint that time one has to
distinguish real trends from false ones due to correlated noise.  This
is always tricky, but a troublesome detail is that at that point the
results tend to have a non-Gaussian distribution,\cite{Hetherington84}
which makes it difficult to produce error bars with a well defined
statistical meaning. In addition, there is the time-step error, which
arrises because the imaginary-time evolution operator $\exp(-t\H)$ has
to be evaluated as the limit $\tau\to 0$ of $[\exp(-\tau\H) +
O(\tau^2)] ^{t/\tau}$, but this error is much easier to control.

Next we present results for excited state energies for clusters with
up to seven atoms.  First, we computed energies for trimers of Ne, Ar,
Kr, and Xe ($m=7.092\times 10^{-3},\ 6.9635\times 10^{-4},\ 
1.9128\times 10^{-4}$, and $7.8508\times 10^{-5}$).  Since our
variational functions contain general three-body correlations, the
accuracy of the wavefunctions and energies for the trimers can be
improved without apparent limit other than the machine precision.
During optimization of the wavefunctions for the trimers, we typically
start with the ground state wavefunction which has prefactor degree of
5 or 6. For the trimers we chose not to vary the polynomial
coefficients in the exponent and simply used the fixed terms required
by the boundary conditions.  The quality of the wavefunctions may be
improved by varying polynomial coefficients in the exponent, and for
larger clusters it becomes important to include such polynomials.

\begin{table}[hb]
\caption{Energy levels $E_k$ of the rare gas trimers; the errors are 
estimated to be a few units in the least significant digit.}
\begin{tabular}{l|l|l|l|l}
$k$&  Ne$_3$  &  Ar$_3$       & Kr$_3$         & Xe$_3$\\
\tableline
1& -1.719~560 & -2.553~289~43 & -2.760~555~34 & -2.845~241~50\\
2& -1.222~83  & -2.250~185~5  & -2.581~239~0  & -2.724~955~8 \\
3& -1.142~0   & -2.126~361    & -2.506~946~8  & -2.675~064~8 \\
4& -1.038     & -1.996~43     & -2.412~444    & -2.608~615   \\
5& -0.890     & -1.946~7      & -2.387~973    & -2.592~226   \\
\end{tabular}
\label{Trimer.table}
\end{table}
For the optimization we used samples consisting of 4000
configurations, and we gradually increased the prefactor degree to
improve the quality of the trial functions.  For Ne trimers we
performed \dMC\cite{UNR93} calculations using optimized wavefunctions
with prefactor degrees up to 14.  Although, in priciple, for trimers
nothing should preclude further improvement, the observed changes were
statistically insignificant in the 12 to 14 degree range.  Table
\ref{Trimer.table} contains results for degree 12.

There was no statistically significant difference between time steps
$\tau=$0.4, 0.2 and 0.1, and thus no noticeable time-step error.  In
the \dMC\ calculations we used 1.3 million \MC\ steps.  For Ar, Kr,
and Xe trimers we found that the quality of the wavefunctions does not
improve beyond the prefactor degree 10.  The results in Table
\ref{Trimer.table} for the three more massive noble gas atoms were
obtained using trial wavefunctions with such polynomials.  Convergence
with respect to the time-step was established by comparing $\tau=$0.8,
0.6, and 0.4.  The number of \MC\ steps is the same as for Ne. Except
for the energy of the fifth level of Ne, which is 0.009 too high, the
results in Table \ref{Trimer.table} agree with, and in some cases
improve upon, those of Leitner \etal\cite{Leitner91}.

In Table \ref{Argon.table} we present results for the energies of the
first five levels of Ar clusters of sizes 4 through 7. Our method
allows one to go beyond 7 atom clusters, but as one can see from Table
\ref{Argon.table} the statistical error increase with system size.  To
obtain more accurate results for larger clusters it would probably be
helpful to include higher order correlations in the wavefunction. In
the calculations for 4 through 7 atom clusters we used a 10 degree
prefactor and an exponent of degree three.  Again, 1.3 million step
\dMC\ results were compared for $\tau=$0.8, 0.6, and 0.4.

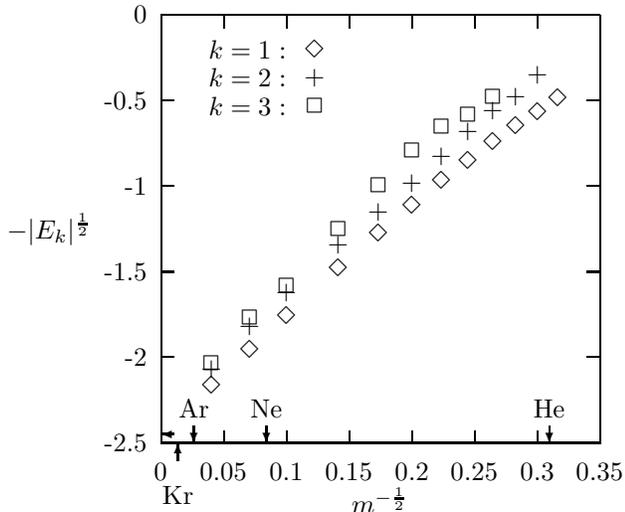
\begin{figure}
\setlength{\unitlength}{0.240900pt}
\ifx\plotpoint\undefined\newsavebox{\plotpoint}\fi
\sbox{\plotpoint}{\rule[-0.200pt]{0.400pt}{0.400pt}}%
\begin{picture}(973,809)(0,0)
\font\gnuplot=cmr10 at 10pt
\gnuplot
\sbox{\plotpoint}{\rule[-0.200pt]{0.400pt}{0.400pt}}%
\put(220.0,113.0){\rule[-0.200pt]{0.400pt}{162.126pt}}
\put(220.0,113.0){\rule[-0.200pt]{4.818pt}{0.400pt}}
\put(198,113){\makebox(0,0)[r]{-2.5}}
\put(889.0,113.0){\rule[-0.200pt]{4.818pt}{0.400pt}}
\put(220.0,248.0){\rule[-0.200pt]{4.818pt}{0.400pt}}
\put(198,248){\makebox(0,0)[r]{-2}}
\put(889.0,248.0){\rule[-0.200pt]{4.818pt}{0.400pt}}
\put(220.0,382.0){\rule[-0.200pt]{4.818pt}{0.400pt}}
\put(198,382){\makebox(0,0)[r]{-1.5}}
\put(889.0,382.0){\rule[-0.200pt]{4.818pt}{0.400pt}}
\put(220.0,517.0){\rule[-0.200pt]{4.818pt}{0.400pt}}
\put(198,517){\makebox(0,0)[r]{-1}}
\put(889.0,517.0){\rule[-0.200pt]{4.818pt}{0.400pt}}
\put(220.0,651.0){\rule[-0.200pt]{4.818pt}{0.400pt}}
\put(198,651){\makebox(0,0)[r]{-0.5}}
\put(889.0,651.0){\rule[-0.200pt]{4.818pt}{0.400pt}}
\put(220.0,786.0){\rule[-0.200pt]{4.818pt}{0.400pt}}
\put(198,786){\makebox(0,0)[r]{0}}
\put(889.0,786.0){\rule[-0.200pt]{4.818pt}{0.400pt}}
\put(220.0,113.0){\rule[-0.200pt]{0.400pt}{4.818pt}}
\put(220,68){\makebox(0,0){0}}
\put(220.0,766.0){\rule[-0.200pt]{0.400pt}{4.818pt}}
\put(318.0,113.0){\rule[-0.200pt]{0.400pt}{4.818pt}}
\put(318,68){\makebox(0,0){0.05}}
\put(318.0,766.0){\rule[-0.200pt]{0.400pt}{4.818pt}}
\put(417.0,113.0){\rule[-0.200pt]{0.400pt}{4.818pt}}
\put(417,68){\makebox(0,0){0.1}}
\put(417.0,766.0){\rule[-0.200pt]{0.400pt}{4.818pt}}
\put(515.0,113.0){\rule[-0.200pt]{0.400pt}{4.818pt}}
\put(515,68){\makebox(0,0){0.15}}
\put(515.0,766.0){\rule[-0.200pt]{0.400pt}{4.818pt}}
\put(614.0,113.0){\rule[-0.200pt]{0.400pt}{4.818pt}}
\put(614,68){\makebox(0,0){0.2}}
\put(614.0,766.0){\rule[-0.200pt]{0.400pt}{4.818pt}}
\put(712.0,113.0){\rule[-0.200pt]{0.400pt}{4.818pt}}
\put(712,68){\makebox(0,0){0.25}}
\put(712.0,766.0){\rule[-0.200pt]{0.400pt}{4.818pt}}
\put(811.0,113.0){\rule[-0.200pt]{0.400pt}{4.818pt}}
\put(811,68){\makebox(0,0){0.3}}
\put(811.0,766.0){\rule[-0.200pt]{0.400pt}{4.818pt}}
\put(909.0,113.0){\rule[-0.200pt]{0.400pt}{4.818pt}}
\put(909,68){\makebox(0,0){0.35}}
\put(909.0,766.0){\rule[-0.200pt]{0.400pt}{4.818pt}}
\put(220.0,113.0){\rule[-0.200pt]{165.980pt}{0.400pt}}
\put(909.0,113.0){\rule[-0.200pt]{0.400pt}{162.126pt}}
\put(220.0,786.0){\rule[-0.200pt]{165.980pt}{0.400pt}}
\put(45,449){\makebox(0,0){$-|E_k|^{1 \over 2}$}}
\put(564,23){\makebox(0,0){$m^{-{1 \over 2}}$}}
\put(247,32){\makebox(0,0){Kr}}
\put(272,167){\makebox(0,0){Ar}}
\put(386,167){\makebox(0,0){Ne}}
\put(830,167){\makebox(0,0){He}}
\put(220.0,113.0){\rule[-0.200pt]{0.400pt}{162.126pt}}
\put(240,127){\vector(-1,0){20}}
\put(247,86){\vector(0,1){27}}
\put(272,140){\vector(0,-1){27}}
\put(386,140){\vector(0,-1){27}}
\put(830,140){\vector(0,-1){27}}
\put(417,732){\makebox(0,0)[r]{$k=1: $}}
\put(461,732){\raisebox{-.8pt}{\makebox(0,0){$\Diamond$}}}
\put(299,203){\raisebox{-.8pt}{\makebox(0,0){$\Diamond$}}}
\put(359,260){\raisebox{-.8pt}{\makebox(0,0){$\Diamond$}}}
\put(417,313){\raisebox{-.8pt}{\makebox(0,0){$\Diamond$}}}
\put(498,388){\raisebox{-.8pt}{\makebox(0,0){$\Diamond$}}}
\put(561,443){\raisebox{-.8pt}{\makebox(0,0){$\Diamond$}}}
\put(614,487){\raisebox{-.8pt}{\makebox(0,0){$\Diamond$}}}
\put(660,525){\raisebox{-.8pt}{\makebox(0,0){$\Diamond$}}}
\put(702,557){\raisebox{-.8pt}{\makebox(0,0){$\Diamond$}}}
\put(741,586){\raisebox{-.8pt}{\makebox(0,0){$\Diamond$}}}
\put(777,611){\raisebox{-.8pt}{\makebox(0,0){$\Diamond$}}}
\put(811,633){\raisebox{-.8pt}{\makebox(0,0){$\Diamond$}}}
\put(843,655){\raisebox{-.8pt}{\makebox(0,0){$\Diamond$}}}
\put(417,687){\makebox(0,0)[r]{$k=2: $}}
\put(461,687){\makebox(0,0){$+$}}
\put(299,229){\makebox(0,0){$+$}}
\put(359,296){\makebox(0,0){$+$}}
\put(417,349){\makebox(0,0){$+$}}
\put(498,425){\makebox(0,0){$+$}}
\put(561,476){\makebox(0,0){$+$}}
\put(614,522){\makebox(0,0){$+$}}
\put(660,564){\makebox(0,0){$+$}}
\put(702,603){\makebox(0,0){$+$}}
\put(741,635){\makebox(0,0){$+$}}
\put(777,657){\makebox(0,0){$+$}}
\put(811,691){\makebox(0,0){$+$}}
\sbox{\plotpoint}{\rule[-0.400pt]{0.800pt}{0.800pt}}%
\put(417,642){\makebox(0,0)[r]{$k=3: $}}
\put(461,642){\raisebox{-.8pt}{\makebox(0,0){$\Box$}}}
\put(299,238){\raisebox{-.8pt}{\makebox(0,0){$\Box$}}}
\put(359,309){\raisebox{-.8pt}{\makebox(0,0){$\Box$}}}
\put(417,360){\raisebox{-.8pt}{\makebox(0,0){$\Box$}}}
\put(498,449){\raisebox{-.8pt}{\makebox(0,0){$\Box$}}}
\put(561,518){\raisebox{-.8pt}{\makebox(0,0){$\Box$}}}
\put(614,572){\raisebox{-.8pt}{\makebox(0,0){$\Box$}}}
\put(660,610){\raisebox{-.8pt}{\makebox(0,0){$\Box$}}}
\put(702,629){\raisebox{-.8pt}{\makebox(0,0){$\Box$}}}
\put(741,657){\raisebox{-.8pt}{\makebox(0,0){$\Box$}}}
\end{picture}
\caption{$-\sqrt{-E_k}$ of lowest three levels ($k=1,2,3$) for four
  atom clusters vs $m^{-{1\over 2}}$.  The estimated errors for most
  energies are smaller, than the plot symbols, and increase for
  decreasing mass.  Missing data points indicate that no reliable
  estimates were obtained.  The vertical arrows indicate Kr, Ar, Ne,
  and He; the horizontal arrow indicates the classical value -$\sqrt
  6$.}
\label{fig.Evsm}
\end{figure}

As to the performance of our method as the mass $m$ decreases and the
atoms become more weakly bound, we find that both the optimization and
the projection methods begin to fail, because the elementary basis
functions lack the required variational freedom.  This breakdown is
illustrated in Fig.~\ref{fig.Evsm}, which contains three energy levels
as a function of mass for a four atom cluster.  The results are
plotted using variables chosen so that there is linear dependence both
\begin{table}[hb]
\caption{Energy levels of Ar clusters of up to seven atoms; the errors are 
estimated to be a few units in the least significant digit.}
\begin{tabular}{l|l|l|l|l}
$k$&  Ar$_4$ & Ar$_5$ & Ar$_6$    & Ar$_7$\\
\tableline
1& -5.118~11 & -7.785~1 & -10.887~9 & -14.191\\
2& -4.785    & -7.567   & -10.561   & -13.969\\
3& -4.674    & -7.501   & -10.51    & -13.80 \\
4& -4.530    & -7.39    & -10.46    & -13.74 \\
5& -4.39     & -7.36    & -10.35    & -13.71 \\
\end{tabular}
\label{Argon.table}
\end{table}
\noindent
for large masses and for energies close to zero.\cite{MMN96} As the
energy of the levels approaches zero, the scatter in the data points
increases, and ultimately the method fails to produce reliable
results.  Again, the use of trial wavefunctions with four-body
correlations is likely to make it possible to continue to smaller
masses.

As we mentioned before, exponential scaling with dimensionality limits
the applicability of the DVR method.  We can only speculate how the
\MC\ method discussed here scales, since the accuracy of the results is
determined mostly by the elusive quality of the basis functions.  It is
precisely the degradation of the trial functions which is responsible
for the big differences in accuracy among the results we presented.
Clearly, more highly excited states have more structure, but the
harmonic approximation suggests that the corresponding increase in
complexity scales with a low power of the excitation level. It is
plausible that it suffices to include in the elementary basis functions
$n$-body correlations upto some finite order $n$ only, which suggest
polynomial complexity for the computation of the basis functions.  Much
evidence suggests that the projection stage of the calculations also
scales with a small power of system size.  We performed the projection
part of the calculations by a variant of pure-diffusion Monte
Carlo\cite{caffarel,CeperleyBernu88}.  The statistical noise of this
approach for large systems increases exponentially, but there are
alternatives to avoid this, and there it is likely that the usual power
law behavior of \dMC\ methods\cite{FoulkesMitasNeedsRajagopal} can be
recovered.  Let it suffice to say that the longer runs typically took a
few hours on a four processor SGI Origin 200, to produce the results
presented here.

This research was supported by the (US) National Science Foundation
(NSF) through Grant DMR-9725080.  It is our pleasure to thank David
Freeman and Cyrus Umrigar for valuable disscusions and suggestions for
improvements of the manuscript.


\begin{references}
\bibitem{close.to.0} We ignored all singular values $\sigma_k$ with
  $\sigma_k < 10^3 \sigma_1 \epsilon_{\rm dbl}$, where
  $\epsilon_{\rm{dbl}}$ is the double precision machine accuracy.
\end{references}
\end{document}